\documentclass{aa}
\usepackage{graphics}
\usepackage{longtable}
\usepackage{txfonts}
\usepackage{graphicx}
\usepackage{hyperref}
\usepackage{multirow}

\begin{document}

\title{Imaging the spinning gas and dust in the disc around the
  \\ supergiant A[e] star HD~62623
\thanks{
  Based on CNRS Guaranteed Time Observations with ESO telescopes at
  the Paranal Observatory under program 084.D-0355, and on
  Director's Discretionary Time, 284.D-5059. Feasibility was
  assessed using open time, 083.C-0621.
}
}

\authorrunning{Millour et al.}
\titlerunning{Images of the spinning disc around HD~62623}

\author{F. Millour  \inst{1,2}, A. Meilland \inst{2}, O. Chesneau \inst{1}, Ph. Stee
  \inst{1}, S. Kanaan \inst{1,3}, R. Petrov
  \inst{1}, D. Mourard \inst{1}, and S.Kraus \inst{2,4}}

\offprints{fmillour@oca.eu}

\institute{
  Laboratoire FIZEAU, Universit\'e de Nice-Sophia Antipolis,
  Observatoire de la C\^ote d'Azur, 06108 Nice, France
  \and
  Max-Planck-Institute for Radioastronomy, Auf dem H\"ugel 69, 53121,
  Bonn, Germany
  \and
  Departamento de F\'isica y Astronom\'ia, Universidad de
  Valpara\'iso, Errázuriz 1834, Valparaíso, Chile.
  \and
  Department of Astronomy, University of Michigan, 500 Church Street,
  Ann Arbor, Michigan 48109-1090, USA
}

\date{Received; accepted }

\abstract
    {
      To progress in the understanding of evolution of massive stars
      one needs to constrain the mass-loss and determine the
      phenomenon responsible for the ejection of matter an its
      reorganization in the circumstellar environment
    }
    {
      In order to test various mass-ejection processes, we probed the
      geometry and kinematics of the dust and gas surrounding the A[e]
      supergiant HD~62623.
    }
    {
      We used the combined high spectral and spatial resolution offered
      by the VLTI/AMBER instrument. Thanks to a new multi-wavelength
      optical/IR interferometry imaging technique, we reconstructed the
      first velocity-resolved images with a milliarcsecond
      resolution in the infrared domain.
    }
    {
      We managed to disentangle the dust and gas emission in the HD~62623
      circumstellar disc. We measured the dusty disc inner inner rim, i.e. 6
      mas, constrained the inclination angle and the position angle of the
      major-axis of the disc. We also measured the inner gaseous disc
      extension (2 mas) and probed its velocity field thanks to
      AMBER high spectral resolution. We find that the expansion
      velocity is negligible, and that Keplerian rotation is a favoured velocity field. Such a velocity field is unexpected if fast rotation of the central star alone is the main mechanism of matter ejection.
    }
    {
      As the star itself seems to rotate below its breakup-up
      velocity, rotation cannot explain the formation of the dense
      equatorial disc. Moreover, as the expansion velocity is
      negligible, radiatively driven wind is also not a suitable
      explanation to explain the disc formation. Consequently, the
      most probable hypothesis is that the accumulation of matter in
      the equatorial plane is due to the presence of the spectroscopic
      low mass companion.
    }
    \keywords{  Techniques: imaging spectroscopy --
      Stars: emission-line, Be  --
      Techniques: interferometric  --
      Stars: individual: HD~62623 --
      Techniques: high angular resolution --
      Stars: circumstellar matter
    }

    \maketitle
    %
    %________________________________________________________________

    \section{Introduction} 

    The supergiant A[e] star (\cite{L98}) HD~62623 (HR~2996,
    3~Puppis, l~Puppis), is a key object for understanding the processes at the
    origin of aspherical shells in massive evolved stars (\cite{HD94})
    and supernovae (\cite{K87}). Indeed, HD~62623 is surrounded by a
    dense gaseous and dusty disc (\cite{M10}), a structure more often
    found in young stellar objects and post-AGB stars
    (\cite{vW06}). Discs are known to govern accretion or mass-loss in
    these lower-mass objects, but their origin and structure remain
    highly debated for massive stars (\cite{P02}).

Fast rotation of the star leads to an expanding disc-like wind in the case of very massive stars (\cite{Z85} ; \cite{LP91}) or, when viscosity becomes dominant in less massive stars, it leads to a rotating disc (\cite{L91}).
The presence of a companion star could also lead to a rotating disc  (\cite{P95}). 

Fast rotation, or the presence of a companion star, could be responsible for the breakup of spherical symmetry of mass-loss around the massive, hot and luminous object at the core of HD~62623. To test these two hypotheses, one needs to access
    unprecedented combined spatial and spectral resolutions. Here, we
    report the first continuum image and velocity-resolved images in a circumstellar disc
    using a new multi-wavelength optical interferometry imaging
    method, which allows us to spatially disentangle the dust and gas
    emissions of HD~62623.

    \begin{table}[htbp]
      {\centering \begin{tabular}{ccccc}

          \hline 
          Date&Telescope&Number&Seeing&Coherence\\
	  &Config.&of Obs &(")& time (ms)\\
          \hline 
          \hline

          08/01/10&D0-H0-K0&10&0.48 -- 1.48&2.5 -- 7.3\\
          11/01/10&D0-G1-H0&1&0.84&4.9\\
          17/01/10&E0-G0-H0&13&0.50 -- 1.03&2.4 -- 5.1\\
          19/01/10&A0-K0-G1&10&0.72 -- 1.42&4.3 -- 8.4\\
          18/03/10&D0-G1-H0&2&0.55 -- 1.68&1.7 -- 2.7\\

          \hline

        \end{tabular}\par}\caption{\label{log}VLTI/AMBER observing log
        for HD62623.}
    \end{table} 

    %________________________________________________________________

    \section{VLTI/AMBER Observations, and Data Reduction}

\begin{figure}[htbp]
      \centering
    \includegraphics[width=0.4\textwidth]{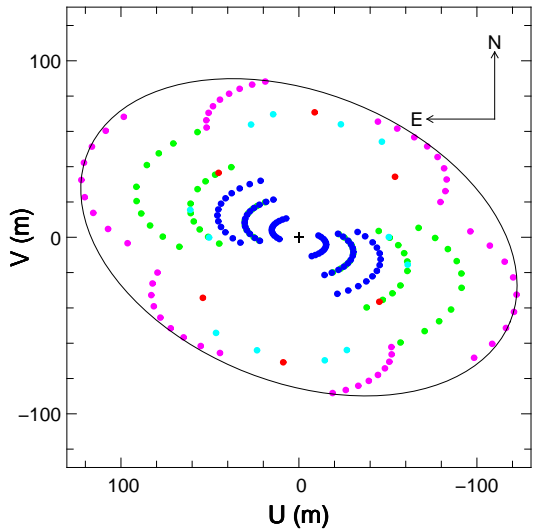}
      \caption{
      The UV coverage of our observations. Different colours represent different nights. The ellipse has dimensions 82x128m.
}
      \label{uv}
    \end{figure}

\begin{figure*}[htbp]
      \centering
    \includegraphics[height=0.325\textwidth, origin=br, angle=-90]{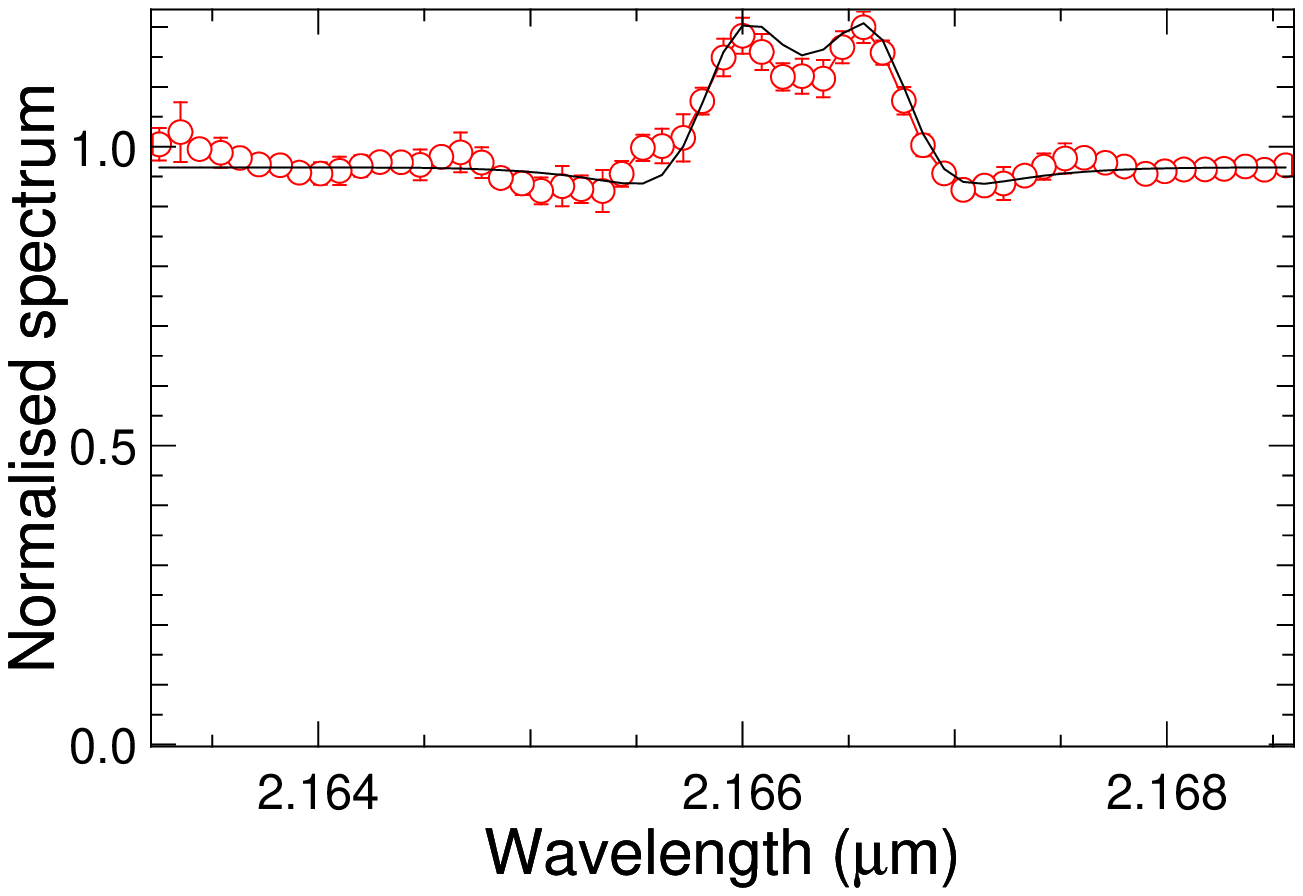}
    \includegraphics[height=0.65\textwidth, origin=br, angle=-90]{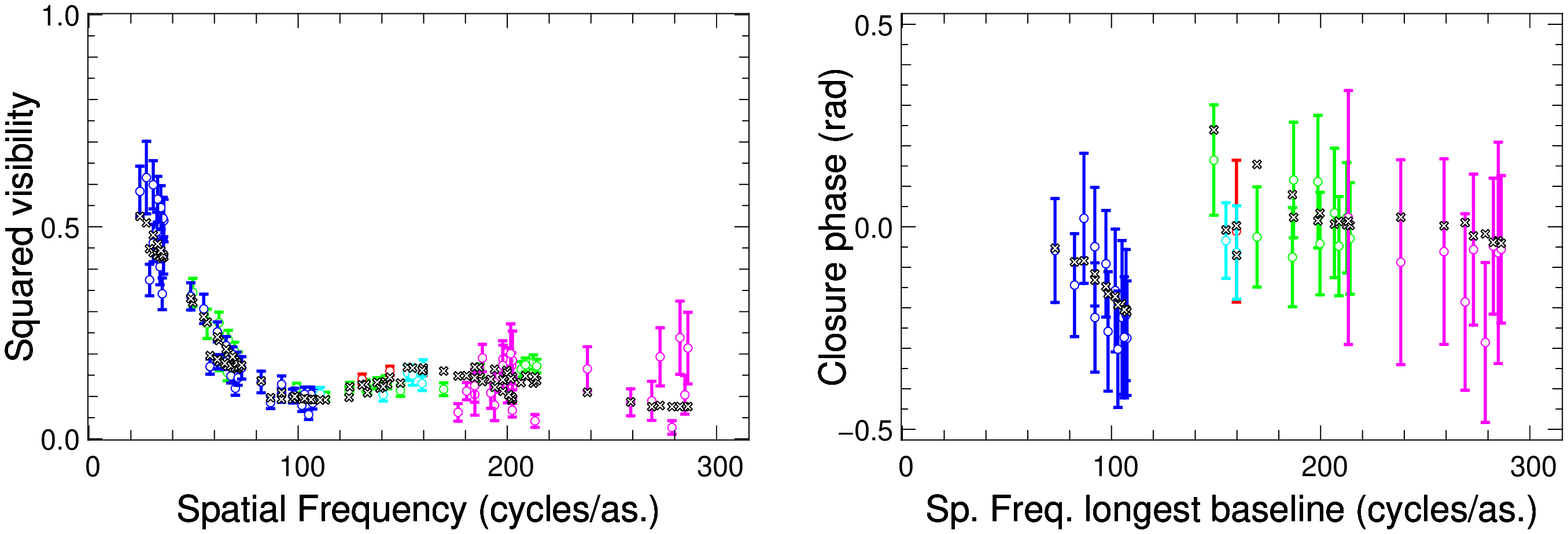}\\
    \includegraphics[height=1.0\textwidth, angle=-90]{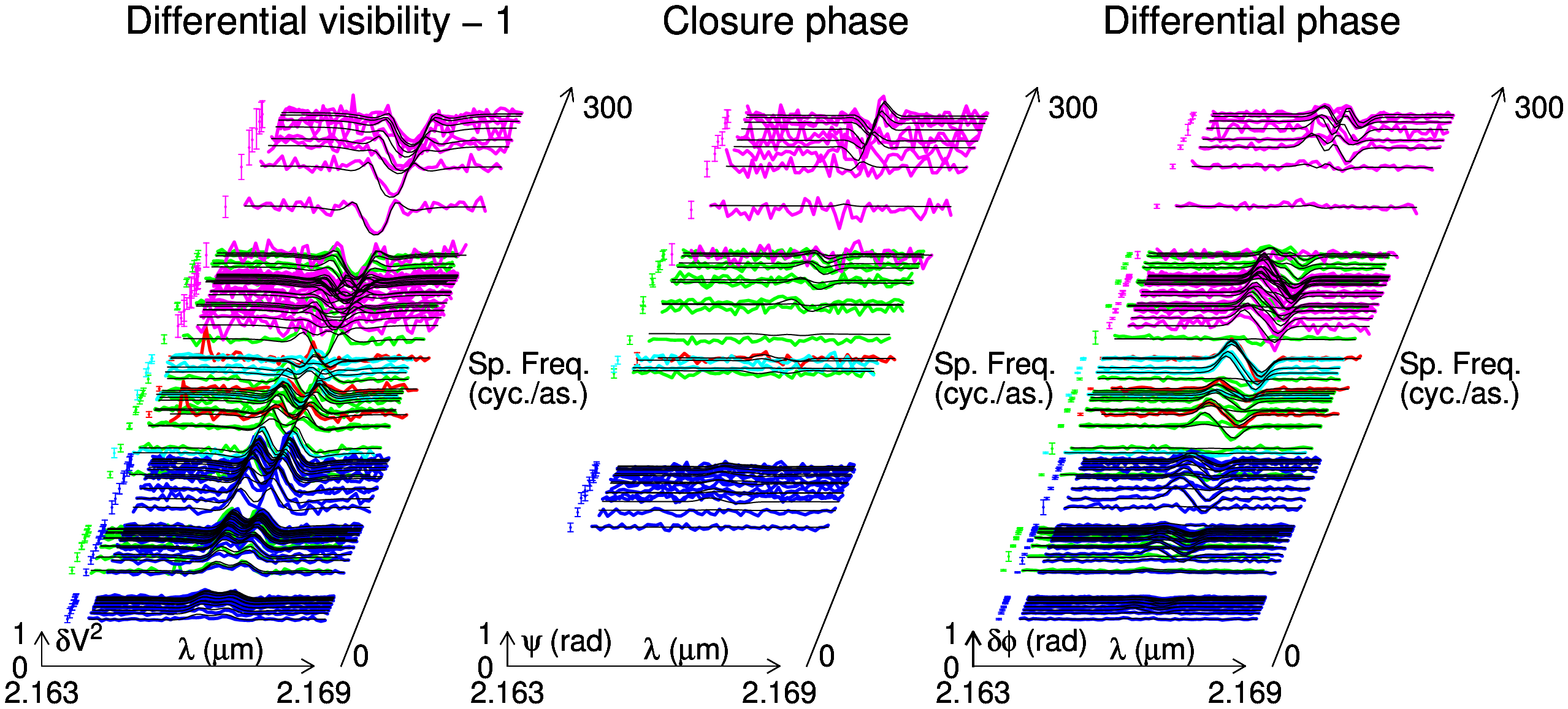}
      \caption{
{\bf Top-Left:} The spectrum. The double-peaked emission line is the Br$\gamma$ line. {\bf Top-Middle:} The squared visibilities as a function of spatial frequency, {\bf Top-Right:} the closure phases as a function of spatial frequency, {\bf Bottom-Left:} the differential visibilities (visibilities normalized to 1 in the continuum), {\bf Bottom-Middle:}~the closure phase, and {\bf Bottom-Right:} the differential phase, the three former as a function of wavelength. In all plots, different colours represent different nights, and the black lines (resp. the black dots for the top panels) represent our best-fit model. "cycles/as." or "cyc./as." means "cycles per arc-seconds" or "arc-seconds$^{-1}$". For the closure phase plot, the spatial frequencies are the ones of the longest baseline. In the squared visibilities panel, one can see that the disc is clearly resolved at the largest baselines.
}
      \label{data}
    \end{figure*}

    To unambiguously resolve the close environment of HD~62623, both
    spatially and spectrally, we acquired data in early 2010
    (table~\ref{log}) using the VLTI/AMBER instrument
    (\cite{P07}). Each AMBER measurement consists of three
    visibilities, one closure phase, three wavelength-differential
    phases, and one flux measurement, each of them spectrally
    dispersed on about 500 narrow-band channels, with a spectral
    resolution power of 12000 close to the Br$\gamma$ line. We used
    the standard AMBER package (\cite{T07}) to reduce the data,
    complemented with calibration scripts (\cite{M08}). The dataset ($\approx54000$ visibilities, $\approx54000$ differential phases, and $\approx18000$ closure phases, presented Fig.~\ref{data}) and resulting UV coverage, presented in Fig~\ref{uv}, are noticeably the largest obtained so far with the VLTI.

%, and
   % saved the result to the standard optical/IR interferometry OI-FITS
    %format (\cite{P05}). The dataset and resulting spatial
    %frequencies coverage presented in Fig~\ref{uv} are noticeably the
    %most complete obtained so far with the VLTI.

Our HD~62623 squared visibilities (top-middle of Fig~\ref{data}) decrease with increasing spatial frequencies with values close to 0 at high frequencies, meaning that we resolved the close environment of HD~62623. In addition, we detect non-zero closure phases, indicating asymmetries in the environment of HD~62623 (top-right panel of Fig.~\ref{data}). In addition, we clearly resolve changes in the visibilities, closure phases and differential phases in the Br$\gamma$ line (bottom panels of Fig.~\ref{data}).

    We then used the MIRA software (\cite{T03}) to recover a spectrally dispersed image cube of HD~62623, using in a first attempt squared visibilities and closure phases. We did not try other image reconstruction software, based on the conclusion of \cite{M09}. In addition, we developed a dedicated self-calibration method (presented in Section~\ref{sect:selfcal}) to add the wavelength-differential phases to the image recovery process, MIRA being used this time with squared visibilities, closure phases and phases. This additional step contribute to link the astrometry of all narrow-band images together, allowing us to measure astrometric changes in the emission line compared to the continuum. 

To calibrate the phases sign (and, hence, the on-sky orientation of the images), we also performed reconstructions using MIRA on the $\theta$ ~Ori~C dataset from \cite{K09}\footnote{ kindly provided by the authors at \texttt{http://www.mpifr-bonn.mpg.de/staff/skraus/files/amber.htm}}

    %________________________________________________________________
    
    \section{The Self-Calibration Method}
\label{sect:selfcal}
    
\subsection{The Wavelength-Differential Phase}
    Wavelength-differential phase in optical/infrared stellar interferometry is a
    partial measure of the object Fourier phase
    $\phi_{\rm object}$($\lambda$). Indeed, the phase measured on a ground-based
    long-baseline interferometer is affected by randomly variable
    perturbations. These perturbations are composed of, by decreasing
    magnitude (in the near-infrared):
\begin{itemize}
\item $\delta$, a variable achromatic Optical Path Difference
    (OPD) between the two telescopes,
\item $\delta_{\rm dry}$, a chromatic dry
    air OPD,
\item $\delta_{\rm wet}$, a chromatic wet air OPD.
\end{itemize}

The measured
    phase $\varphi_{\rm measured}$ takes therefore the following form:

    \begin{equation}
      \varphi_{\rm measured}(t,\lambda) =  \phi_{\rm object}(\lambda)
      + \frac{2 \pi [\delta(t) + \delta_{\rm dry}(t,\lambda) +
          \delta_{\rm wet}(t,\lambda)]} { \lambda}
    \end{equation}

    The differential phase computation algorithm (see \cite{M06} for
    the application to AMBER) aims at removing the term $2 \pi
    [\delta(t) + \delta_{\rm dry}(t,\lambda) + \delta_{\rm wet}(t,\lambda)] /
    \lambda$ from this equation. This term can be described by the
    following first-order Taylor-expansion:

    \begin{equation}
      \frac{2 \pi [\delta(t) + \delta_{\rm dry}(t,\lambda) +
          \delta_{\rm wet}(t,\lambda)]} { \lambda} = \alpha(t) +
      \beta(t)/\lambda + ...
    \end{equation}
    
    A very simple model-fitting to the observed real-time phase data
    yield $\alpha$ and $\beta$ time sequences. A subsequent
    subtraction and averaging is then performed to compute the
    wavelength-differential phase. In this fitting procedure, any
    phase term from the object that could mimic an atmospheric OPD
    will be removed. Therefore, at first order, differential phase is
    the object's phase where both an offset and a slope have been
    subtracted: 

    \begin{equation}
      \varphi_{\rm diff}(\lambda) = \phi_{\rm object}(\lambda) -
      \alpha' - \beta'/\lambda
    \end{equation}

    In practice, removing this offset and slope from the object phase
    makes the wavelength-differential phase free of any absolute
    astrometric information from the object. On the other hand, the
    wavelength-differential phase does give relative astrometry
    information at one wavelength compared to the one at another
    wavelength, giving access to spectro-astrometric measurements (see \cite{Mi07} for a detailed example of application).

\subsection{Self-Calibration applied to Optical/Infrared Interferometry}
    Compared to the radio-astronomy phase, this optical/infrared
    wavelength-differential phase bears many similarities. The biggest
    difference is that its average is set to zero and that its average
    slope is also set to zero. A previous work (\cite{Mi06}, pp. 63--69, especially Fig.~3.7) demonstrated, in theory, the potential of wavelength-differential  phase for use in interferometric image synthesis. Inspired by a radio-astronomy review article by \cite{PR84}, we realise here the practical application of this development, by implementing an equivalent
    to the Hybrid Mapping algorithm (also called self-calibration
    algorithm), using the MIRA software instead of the CLEAN algorithm
    to produce the hybrid maps, and the wavelength-differential phases
    instead of radio-phases. The main differences in our case compared to
    radio-interferometry are:

    \begin{itemize}
    \item the measured phases are adjusted with a linear-fit to the
      trial ones instead of just offset. This is due to the way the
      wavelength-differential phase is obtained.
    \item the first image-reconstruction step is made using only
      squared visibilities and closure phases, and each narrow-band
      image is centred to its brightest point.
    \end{itemize}

\begin{figure*}[htbp]
      \centering  
      \includegraphics[width=1.0\textwidth]{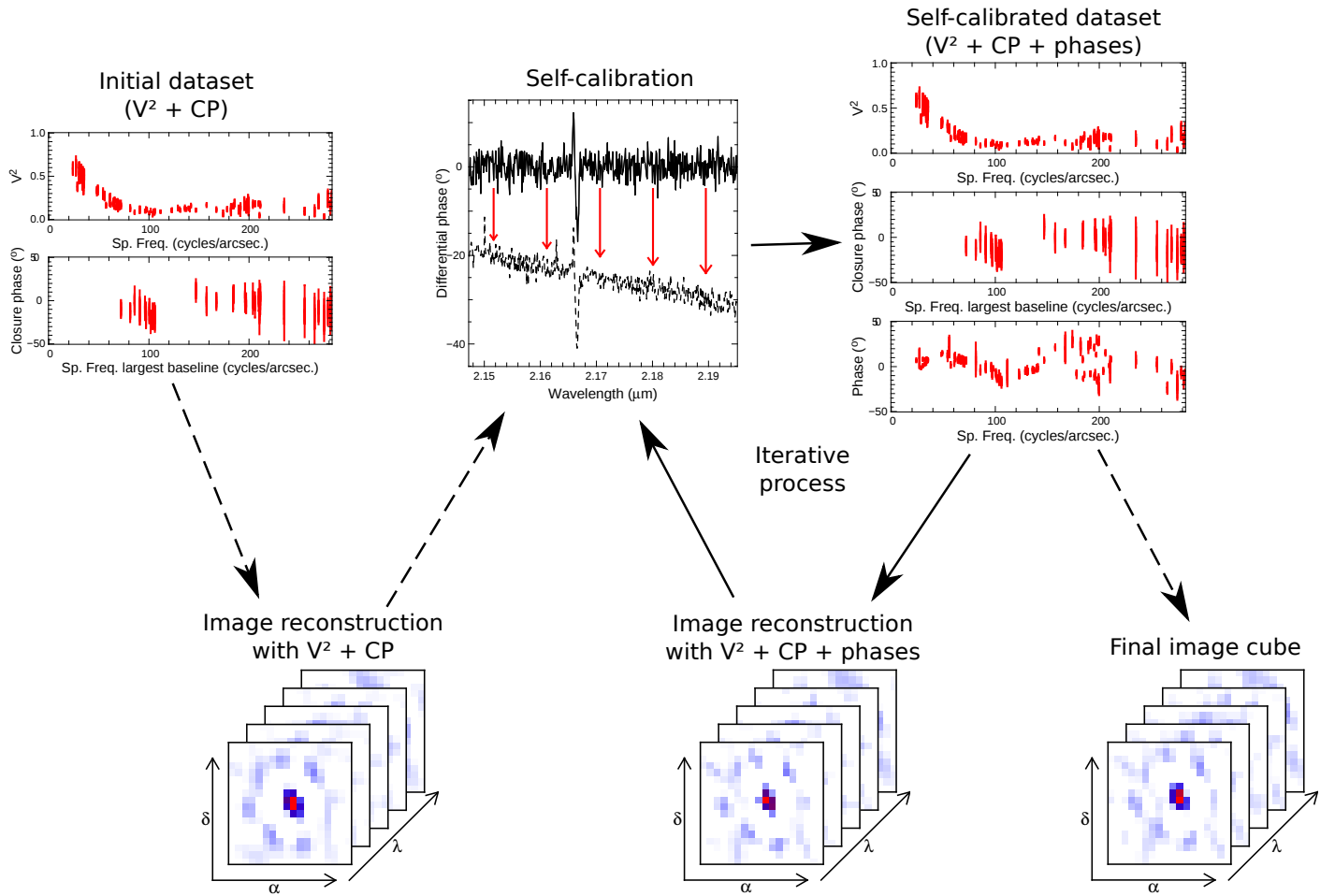}
      \caption{The principle of the self-calibration algorithm applied
        to optical/IR interferometry. Solid arrows show processes that
        may be repeated several times, while dashed arrows show
        processes carried out once.}
      \label{selfcal}
    \end{figure*}

    A sketch explaining this method is shown in Fig~\ref{selfcal}, and a summary of the procedure is described below:
\begin{enumerate}
\item We start the image reconstruction
        process for each wavelength from a set of squared visibilities
        and closure phases.
\item The images are centered on their brightest
        point to limit the astrometric changes between these
        independent images.
\item Then, the self-calibration process itself
        is applied, by comparing the observed wavelength-differential
        phases (solid line in the middle-top panel of Fig.~\ref{selfcal}) to the phase
        computed from the image cube (hybrid map, dashed line in Fig.~\ref{selfcal}). The
        whole dataset is modified so that both the phase offset and
        slope match the image cube ones.
\item Then, the image
        reconstruction is performed again, adding the synthetic phases
        data to the squared visibilities and closure phases.
\item The
        result of this image reconstruction step can then be input
        again into the self-calibration process (step 3.) or used as the final
        image cube.
\end{enumerate}

In our case, we used two of these iterative steps to reach the final result. Using more iteration does not improve image quality, and at some point introduces images artifacts, like would do any other image-recovery process.

    In complement to the described self-calibration algorithm, we used the following parameters in MIRA for the image reconstruction: a quadratic regularization, and a Gaussian prior (to limit the effective reconstruction field of view to approximately 20-30 mas). These parameters set less constraints on the image reconstruction process than in other works, where a severe limitation of the reconstruction field of view (\cite{M07} ; \cite{Z09}) or assumptions on the circular-symmetry of the object (\cite{L09} ; \cite{C10}) were applied to reconstruct the images.

In the last processing steps, we are averaging all the continuum images to
    produce the continuum map and subtract the continuum image from
    each spectral channel images to produce the emission-line
    images. We tried the overall process using random initial images
    to check the unicity of our solution, and checked also on
    simulated datasets the consistency of the whole image
    reconstruction process.

    %________________________________________________________________

    \section{Results}

\begin{figure*}[htbp]
      \centering  
      \begin{tabular}{cc}
{\bf OBSERVATION} & {\bf MODEL}\\
      \includegraphics[width=0.48\textwidth]{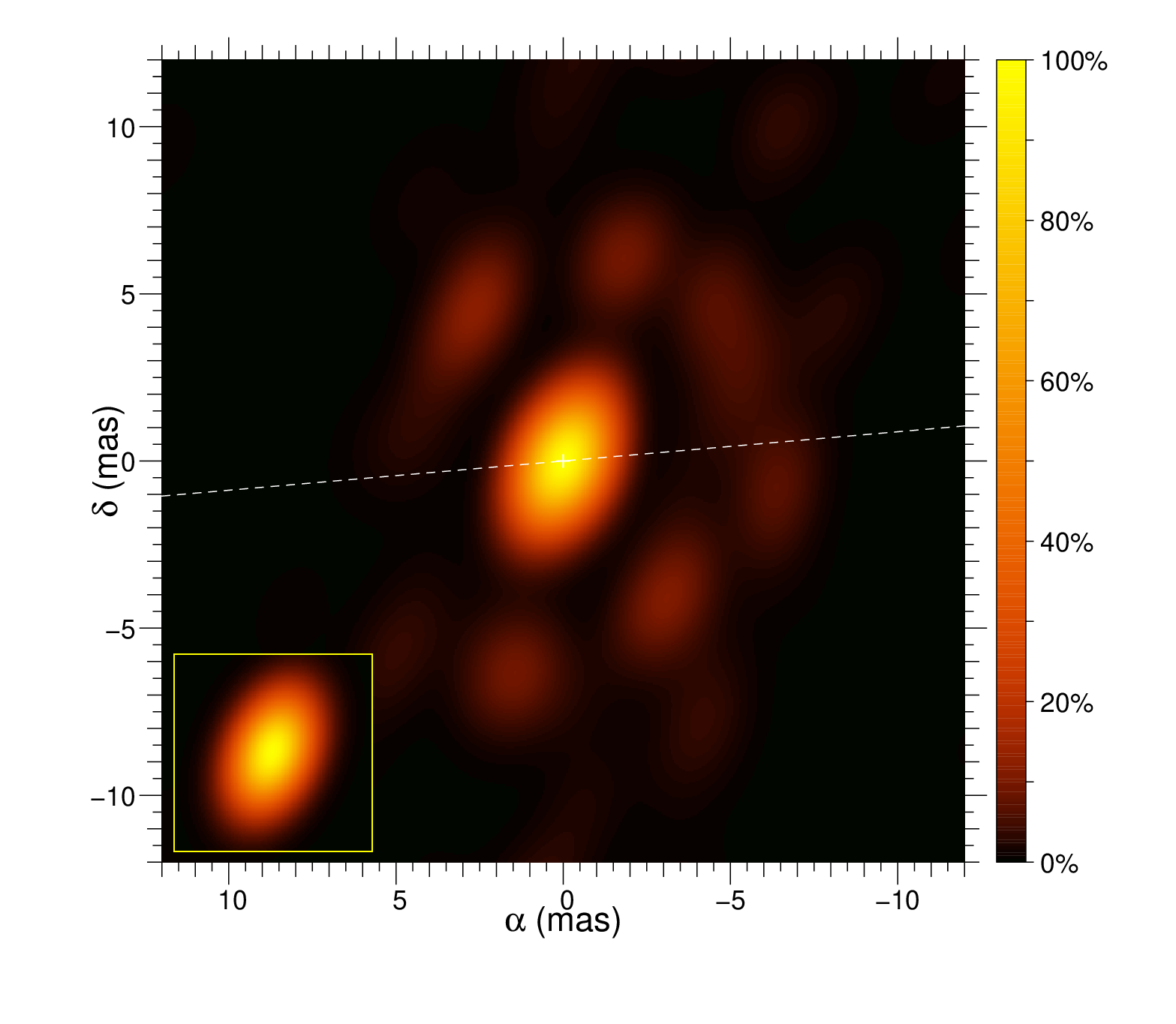}&
      \includegraphics[width=0.48\textwidth]{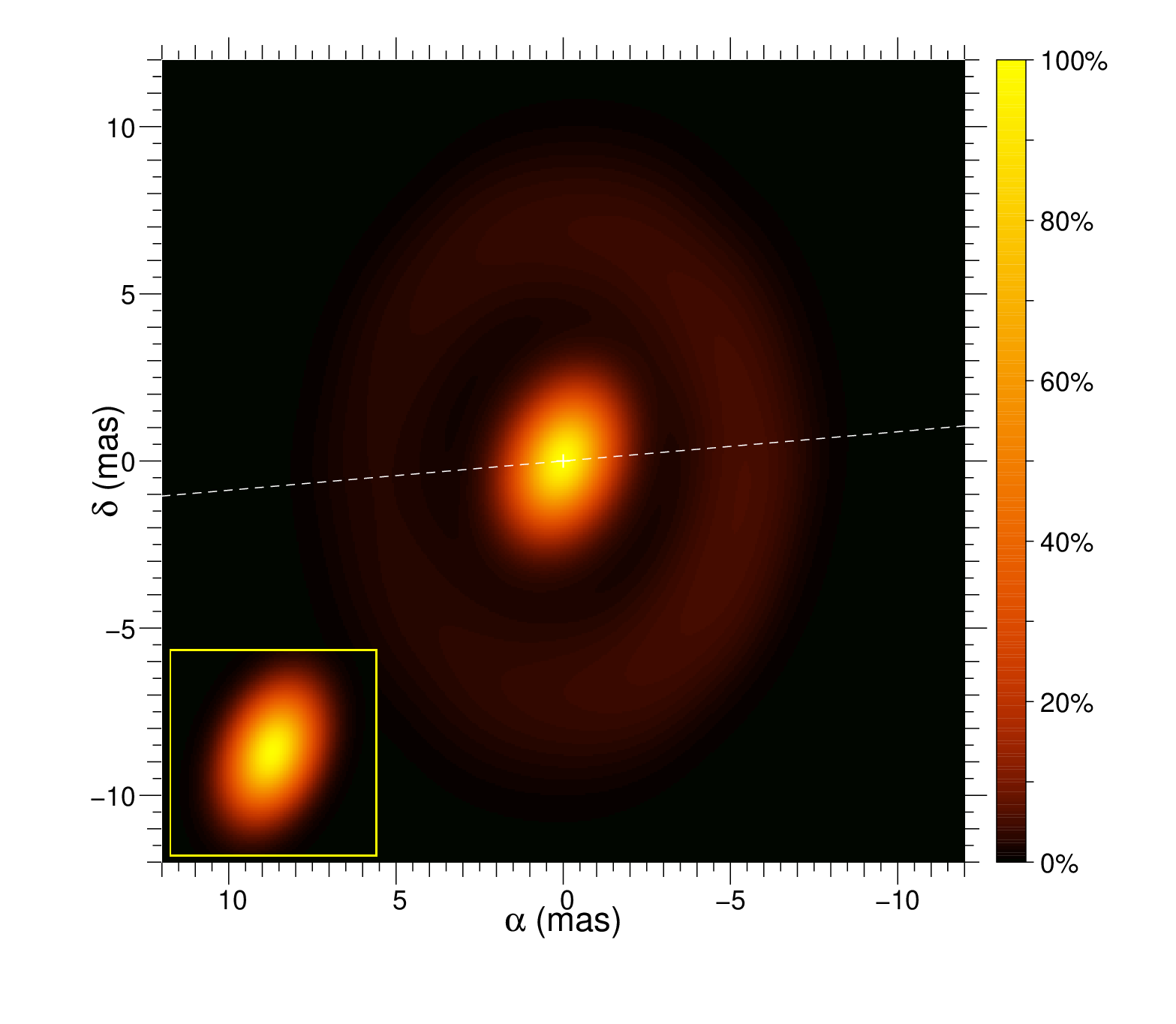}\\
      \includegraphics[width=0.48\textwidth]{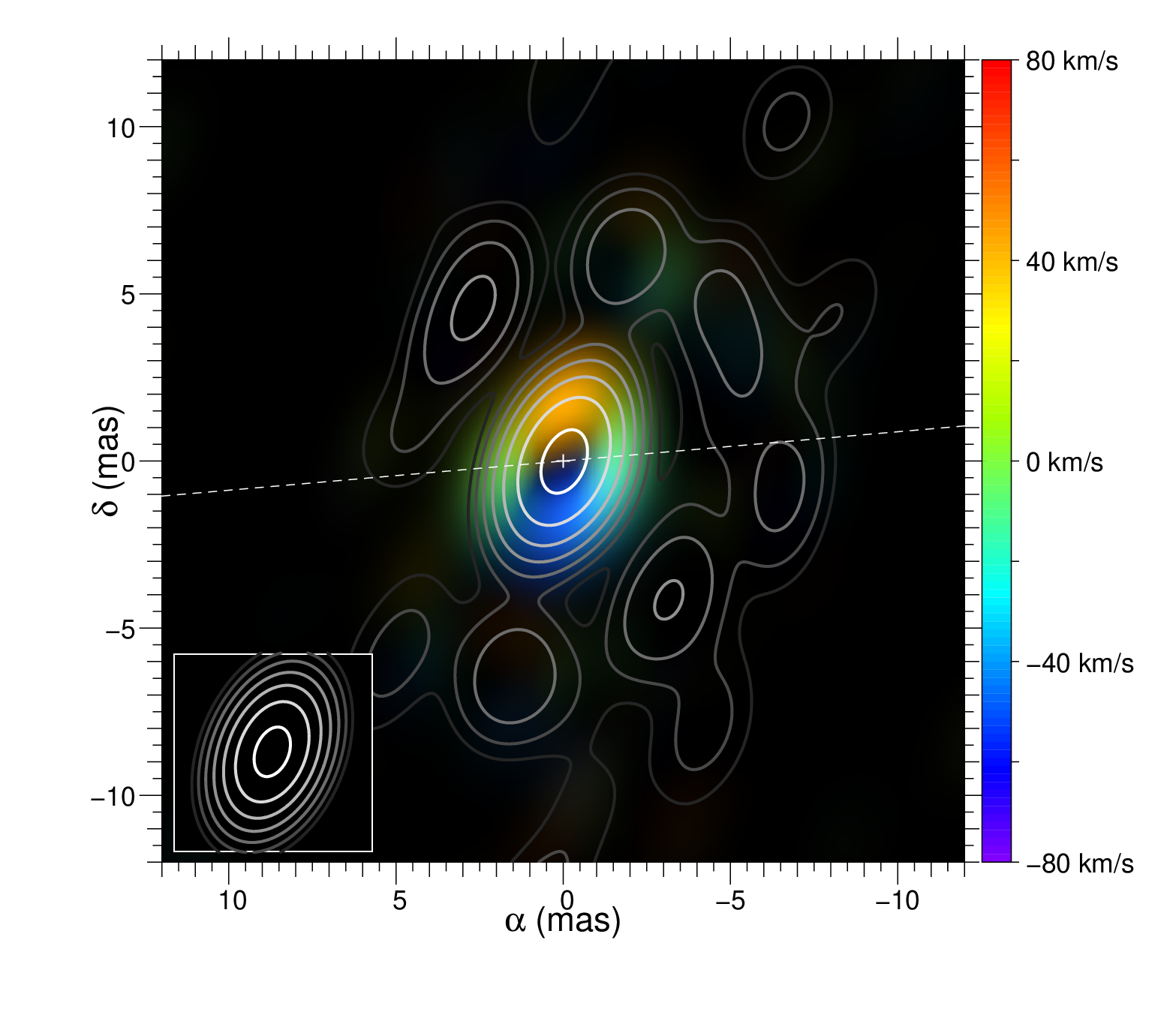}&
      \includegraphics[width=0.48\textwidth]{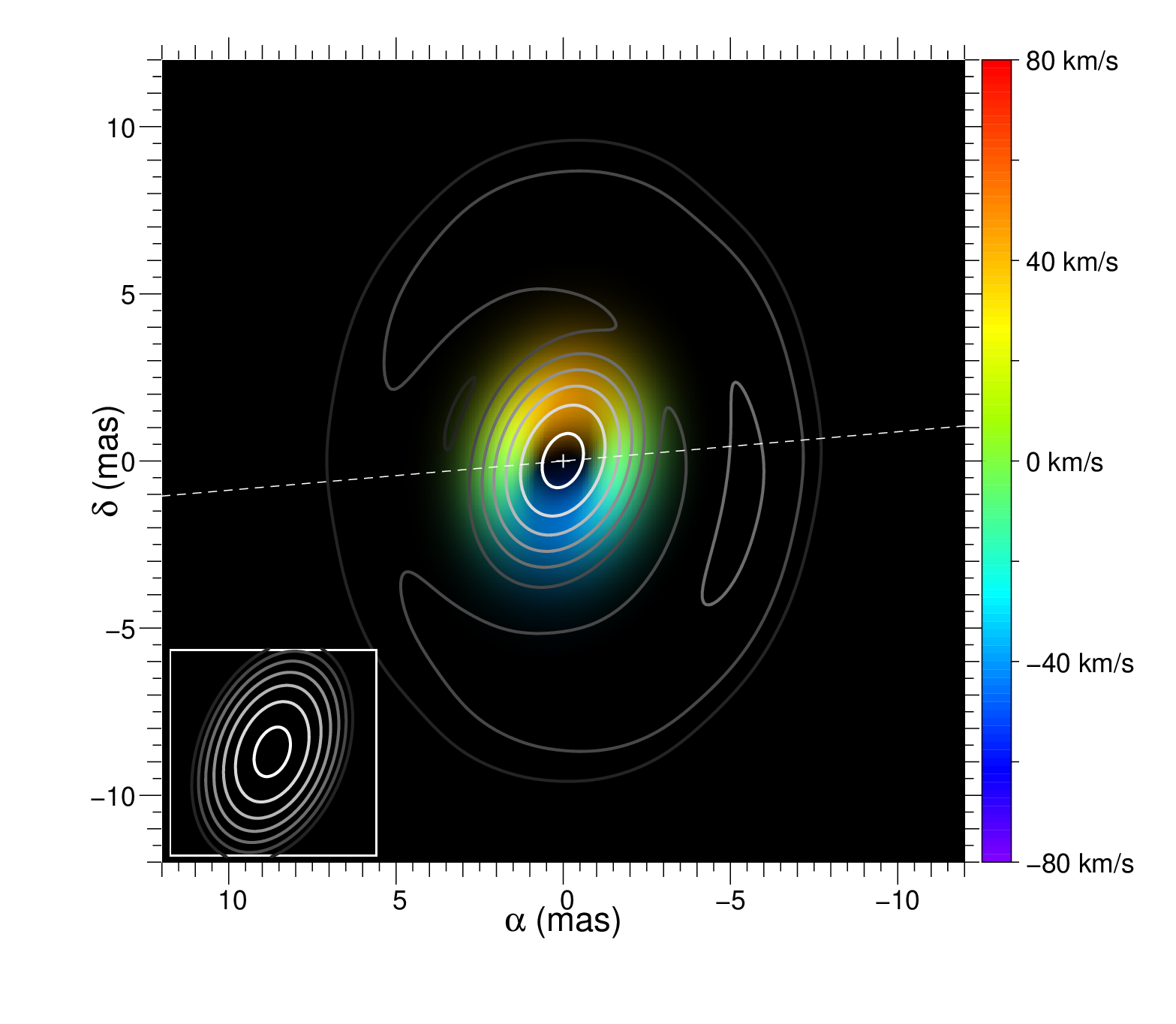}\\
\end{tabular}
      \caption{
{\bf Top-Left:} Composite aperture-synthesis image of HD~62623. The convolving beam of 1.75x2.76 mas is shown in the lower-left box at the same scale. The dashed line represent the polarization measurement from \cite{Y98}. {\bf Top-Right:} Our best-fit model plot at the same scale and convolved with the same beam. {\bf Bottom-left:} Line images of HD~62623. The line intensity is plot as colours for different gas velocities. The intensity in the continuum is plot as gray contours (80\%, 40\%, 20\%, 10\%, 5\%, 2.5\% and 1.25\% of the peak intensity). {\bf Bottom-Right:} Our best-fit model line images with the same representation and the same convolving beam.} 
      \label{image}
    \end{figure*}

    Figure~\ref{image} shows the obtained model-free image where we
    clearly see a ring of material of radius $\approx$6 mas (4 AU at 650 pc)
    encircling a bright inner spot of radius $\approx$2 mas (1.3 AU at 650
    pc). The ring corresponds to the region of the previously-detected
   dusty disc by \cite{M10} where dust is heated at the
    sublimation temperature (inner rim), and the bright spot is a joint
    contribution from the star, and free-free and free-bound emission
    from an inner gaseous disc.

    The continuum image does not show a clear elongation of the ring
    because of the elongated beam in the N-S direction, but clearly
    shows that the ring is brighter on the West side, compared to the
    East side. The inclination angle of the dust disc must therefore
    be intermediate (30 to 60$^o$) and the position angle of the major
    axis of the disc must be close to zero degrees, as we see a
    brighter emission zone opposite to the star and a dimmer, hidden
    emission zone closer to us. This is compatible with the values
    predicted from mid-IR observations (position angle of
    10$\pm$10$^o$, inclination of 60$\pm$10$^o$) and also with the
    measured polarization angle by \cite{Y98}:
    95$\pm$5$^o$, perpendicular to the major axis of the disc.

    Confirmation that the inner spot is partly composed of rotating
    ionized hydrogen (observed here in Br$\gamma$) comes from the line
    emission images, where we see the red-shifted emission appearing
    on the North side of the star, and the blue-shifted emission on
    the South side. At the line center, the emission is roughly
    symmetric and the elongation is directed to the East--West. This
    signal is a typical signature for a rotating disc. The Br$\gamma$
    emission in the disc vanishes rapidly with increasing radius. We
    argue that this is not only a consequence of the decreasing gas
    density, but also of the low ($<$ 10000 K) star temperature, and,
    thus, of the larger neutral hydrogen fraction with increasing
    radius. No Br$\gamma$ emission
    comes from the ring, which means the gaseous hydrogen mixed with
    the dust is not ionized in this region.

    The central star is unresolved by the VLTI in the continuum image,
    but we resolve the central gaseous disc. The lack of line emission
    in the very central region (see Fig.~\ref{lineFlux} and Sect.~\ref{discussion}) implies that the VLTI resolved an
    apparent ``cavity'' (radius not exceeding 1mas, i.e 0.65 AU)
    between the core regions including the A3Ia photospheric stellar
    source (producing an absorption component) and the emitting disc
    of plasma. This confirms that the star's wind do not contribute
    significantly to the Br$\gamma$ emission, and provides an upper limit for
    its radius in accordance to its spectral type, i.e. 65-80
    R$_\odot$ (for d=650 pc, 0.3 AU).

    %________________________________________________________________

    \section{Modelling}

    To match our dataset in the continuum, we used a simple
    geometrical model composed of four components: the star (described
    by an uniform disc with an absorption spectrum), the free-free and
    free-bound emission from the gas (described by an elliptical
    Gaussian disc), the inner rim of the dusty disc (described by an
    elliptical skewed ring) and the outer regions of the disc
    (described here by an extended background).

    In addition, to match the data in the emission-line region, we add
    an elliptical Gaussian-brightness component which is assigned the
    following combination of velocity fields (see \cite{S95} ;
    \cite{Me07} for typical applications of these velocity fields):
    \begin{itemize}
    \item a radial expansion velocity field of the form: $$v_r(r) =
      v_r(0) + [v_r(\infty) - v_r(0) ] [1 - R_* / r]^\gamma$$ with
      $v_r(0)$ the  expansion speed at the smallest radius,
      $v_r(\infty)$ the expansion speed at infinite radius, and
      $\gamma$ the exponent of the power law.
    \item a rotation velocity field of the form: $$v_\theta(r) =
      v_\theta(0) \times r^\beta$$ with $v_\theta(0)$ the maximum
      rotation speed of the disc. When $\beta$ equals -0.5, the disc
      is in Keplerian rotation, when $\beta$ equals 1 the disc is in
      solid rotation, and when $\beta$ equals -1, the disc has a
      constant angular momentum.
    \end{itemize}

Parameters of the whole model range from angular sizes of the components (star, gas disc and dust disc), the system inclination angle, flux ratios between the star, the gas disc, and the dust disc, and also the rotation and expansion velocity, the exponents of the rotation and expansion law, in addition to the on-sky orientation angle.

    This model produces intensity maps that are Fourier-transformed,
    and used to produce synthetic observables (squared visibilities,
    closure phases and differential phases). These synthetic
    observables are adjusted to the observed data using a combination
    of a gradient descent algorithm and a simulated annealing
    algorithm. Given the large amount of observed data, we fitted this model having all parameters potentially free.

In addition to our global fit, we tested several typical models (i.e. fixing few parameters and leaving the other ones free in the fitting procedure described above), including: a model with free rotation and/or expansion laws, a pure Keplerian rotation law, and a solid rotation.

Our model-fitting shows that there is no or very little expansion in the disc around HD~62623, and that rotation is indeed at the core of the disc velocity law.
To check that our best obtained model is located in an absolute minimum, we produced a $\chi^2$ curve, starting from our best solution, fixing the $\beta$ parameter to several values, and fitting the model to the data by using the method described above. 

\begin{figure}[htbp]
      \centering  
      \includegraphics[width=0.45\textwidth]{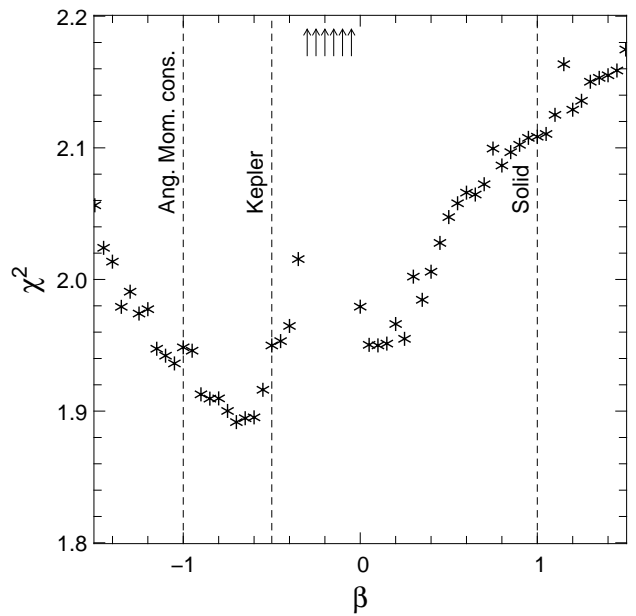}
      \caption{
$\chi^2$ curve obtained by fixing the $\beta$ parameter to different values, and fitting the model to the data. Dashed vertical lines indicate several canonical rotation laws, and arrows represent $\chi^2$ values above 2.2.
	}
      \label{chi2curve}
    \end{figure}

Our final reduced $\chi^2$ is equal to 1.92 for a Keplerian-rotating disc model (See Fig.~\ref{chi2curve}). We find a slightly lower $\chi^2$ for $\beta=-0.7$, i.e. a rotation law between Keplerian rotation and Angular momentum conservation (see Table~\ref{bestparams}). However, all $\chi^2$ values, except for $\beta\approx0$, lie between 1.9 and 2.2. This basically means that the significance of this result is quite low, probably due to the little number of angular-resolution elements we have in the gas disc.

    \begin{table}[htbp]
      {\centering \begin{tabular}{llccc}
          \hline 
          Parameter & Unit & Value \\
          \hline 
          \hline
Flux$_{*}$  & \% total flux  & 34 \\ %3.06 \\
$R_{*}$ & mas & 0.64 \\
Aborption$_{*}$ & \% total flux & 0.23 \\
Absorption$_{*}$ width & $\mu$m & 0.0006 \\
          \hline
Continuum Flux$_{\rm gas\ disc}$ & \% total flux  & 11 \\ %1.0 \\
Continuum FWHM$_{\rm gas\ disc}$ & mas &  1.39\\
Line Flux$_{\rm gas\ disc}$  & \% total flux  & 25\\%8.73 \\
Line FWHM$_{\rm gas\ disc}$ & mas &  1.16\\
$\beta$ & - & -0.7\\
$v_\theta(0)$ & km\textperiodcentered s$^{-1}$ & 221 \\
$\gamma$ & - & 0.86\\
$v_r(0)$ & km\textperiodcentered s$^{-1}$ & 0 \\
$v_r(\infty)$ & km\textperiodcentered s$^{-1}$ & 0\\
\hline
Flux$_{\rm dust\ rim}$ & \% total flux  & 35 \\%3.09 \\
FWHM$_{\rm dust\ rim}$ & mas & 13.9 \\
Thickness$_{\rm dust\ rim}$ & mas & 4.9 \\
Flux$_{\rm extended\ disc}$ &  \% total flux  & 20 \\ %1.78 \\
\hline
Disc inclination & $^o$ & 38 \\
On-sky PA  & $^o$ & 0 \\
          \hline
        \end{tabular}\par}\caption{\label{bestparams}
Best parameters found for the AMBER HD~62623 dataset.
}
    \end{table} 

We conclude that, though the lowest $\chi^2$ values lie close to Keplerian rotation (i.e., rotation law exponent -0.5), our observation only weakly contrain the rotation law of the gas disc. On the other hand, we can clearly rule-out expansion-dominated velocity fields ($\chi^2 > 4$ for all models), confirming that rotation is indeed the main velocity field in the HD~62623 disc.

%    \begin{table*}[htbp]
%      {
%        \centering
%        \begin{tabular}{ccccccccc}
%
%          \hline 
%          Model & $\beta$ & $v_\theta(0)$ & $\gamma$ & $v_r(0)$ & $v_r(\infty)$ & $\chi^2$ \\
%                &         & (km/s)       &          & (km/s)   &
%          (km/s) & \\
%          \hline 
%          \hline
%          
%          Rotation + expansion          & -0.5 & 306 & 0.9   & 0.01 &
%          1.91 \\ 
%          Rotation only                 & -0.5 & 300 & 0     & 0    &
%          1.92 \\
%          Keplerian rotation            & -0.5 & 304 & 0     & 0    &
%          1.92 \\
%          Angular momentum conservation & -1   & 284 & 0     & 0    &
%          2.05 \\
%          Expansion only                &  0   & 0   & 0.003 & 29   &
%          3.89 \\
%          Solid rotation                &  1   & 55  & 0     & 0    &
%          7.83 \\
%
%          \hline
%          
%        \end{tabular}\par}
%      \caption{
%        \label{chi2}
%        All the tested models are sorted by increasing $\chi^2$. The
%        model that best match our dataset is a model including a
%        Keplerian rotating disc ($\beta = -0.5$).
%      }
%    \end{table*} 

    %________________________________________________________________

    \section{Discussion}
\label{discussion}

The spectrally and spatially resolved AMBER/VLTI observations in the continuum and Br$\gamma$ line of HD62623 bring unprecedented information on the disc surrounding an A supergiant.

\begin{figure}[htbp]
      \centering  
      \includegraphics[width=0.48\textwidth]{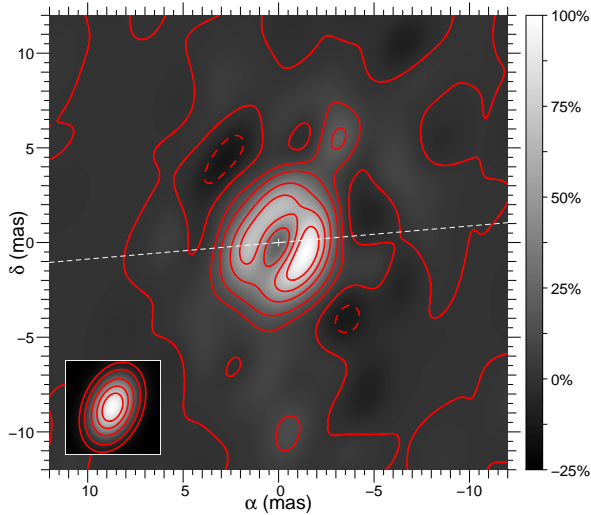}
      \caption{
Line-integrated flux from HD~62623, with the same scale and convolution as Fig~\ref{image}. Solid contours are at 80, 60, 40, 20, 10 and 0\% of the peak intensity. Dashed contours represent negative contributions. Note the doughnut-like shape of the line-emission.
	}
      \label{lineFlux}
    \end{figure}

The central star appears unresolved by the VLTI in the continuum image, providing an upper limit for its radius in accordance for its spectral type, i.e. 65-80\,R$_\odot$ (resp. 130-160) assuming d=650\,pc (resp. 1300\,pc) solar radii. 
The decrease of the line emission in the very central region, giving a doughnut-like shape for the integrated line flux (See Fig.~\ref{lineFlux}), implies that the VLTI resolved an apparent cavity in the innermost region of the emitting disc of plasma (radius not exceeding 1mas, i.e 0.65\,AU), which includes the A3Ia photospheric stellar source. This implies that there is not enough emission in the vicinity of the central star to fill-in this lack of flux. This is a new and clear evidence that the star's wind do not contribute significantly to the Br$\gamma$ emission, and that the spectral type of the central source is not at odd with a normal member of its class. The disc of plasma is therefore at the origin of most, if not all of the emission lines seen the spectrum of HD62623.

The best conditions in terms of temperature and density to generate a Br$\gamma$ flux are met close to the star, and the Br$\gamma$ emission disappears rapidly with increasing radius. We can argue that this is not a only a consequence of the decreasing density, but also of the low (\textless10000k) star temperature and thus the relatively larger neutral hydrogen fraction with 
increasing radius.

    How was such a dense disc formed? 
Even in the case the central
    star would be considered a fast rotator, its rotation rate is lower
    than the nearby rotating plasma ($V \sin i$ ranging from 35 to 80
    km.s$^{-1}$, depending on the work : see for instance \cite{AM95},
    \cite{L72}, \cite{H53}, \cite{W33}).

Previous work by \cite{P95} provided evidence for a companion star to HD~62623, from radial velocity variations. In our images, no companion is detected to about 1\% of the peak intensity (i.e. the typical flux level of image artifacts), an upper limit compatible with the low mass ratio inferred from the Doppler signal (mass of the companion of about 1-2 M$_\odot$).
The angular momentum
    transfer from an orbiting companion may explain a Keplerian
    velocity field. Such an energetic
    argument is widely used in the context of the lower mass evolved
    stars (\cite{vW03}). Furthermore, HD~62623 is a relatively
    cool source among massive stars, exhibiting a radiative wind not
    strong enough to easily justify the accumulation of such dense
    material in the orbital plane in its vicinity.

The spectroscopic observations of \cite{P95} provide an orbital period of 131d or 161d. In the light of the VLTI maps, two options can be considered:
\begin{itemize}
\item one can place the companion close to the supergiant, in the detected cavity, and the plasma and dusty emission originates from a circumbinary disc,
\item one can place the companion further away, perhaps close to the inner rim of the dusty disc. In this case, the disc of plasma is circumprimary, whereas the dust disc can be considered as circumbinary.
\end{itemize}

On the basis of the third Kepler's law, one can exclude a separation larger than 3 AU, since it would imply a total mass of the system larger than 100 Msun (for a period of 161 days). For the same reasons one can also exclude a separation much smaller than 1.5 AU, since it would imply a total mass smaller than few solar masses, i.e smaller than the current mass of the primary assuming a supergiant luminosity class. As demonstrated in \cite{P95}, the mass-ratio is extreme and the primary harbors most of the mass. Hence, the probable separation has to be found within the smallest separations, in the range 1.2-1.7\,AU.

Furthermore, assuming a circular orbit would imply a minimum velocity of about 250-350km/s, not compatible with any observational features detected to date. In particular, the low excitation [OI] forbidden line formed close to the dust forming region shows a double peak with a separation of 55 km/s, suggesting a continuity of the Keplerian kinematics from the dense regions probed by H$\alpha$ or Br$\gamma$ toward the dusty extended ones. Furthermore, significant disturbances of the circumstellar medium would be expected that are not seen in the reconstructed images (which are, however, affected by a limited dynamic range).

We favor the scenario proposed in \cite{P95}, that puts the companion very close to the star, within the Br$\gamma$ emission-free cavity observed in our reconstructed images. An orbital radius of 1.2-1.7 AU corresponding to a total mass of the system of about 15-20 solar masses corresponds at a distance of 650\,pc (resp. 1300\,pc) to 1.8-2.6\,mas (resp. 0.9-1.3), and an efficient transfer of angular momentum route may occur via the L2 co-orbiting Lagrange point, moving at a velocity of 130-170\,km/s. Accounting for the new constraints on the system inclination, this value is in accordance to the H$\alpha$ and Br$\gamma$ velocities of 120\,km/s inferred from the double peaks separation in the emission line, both transitions emitting efficiently from this location. This scenario also provides a natural explanation for the anomalously large mass-loss of the A3 supergiant, whose external layers should experience strong tides under the influence of the close companion.

    %________________________________________________________________
    \section{Conclusion}

    These observations represent a new step in our understanding of
    the formation of flat, dense, Keplerian discs around evolved
    stars. These excretion discs share many physical characteristics with the
    accretion discs encountered around young stellar sources, and are
    now also unambiguously detected around binary post-AGBs,
    i.e. systems in which a low to intermediate mass star expels its
    remaining envelope, in the process of becoming a hot white dwarf
    (\cite{vW03}, \cite{vW06}). In these low-mass binary systems, an efficient
    mass-transfer has occurred when one of the star gets giant,
    leading to the formation of potentially long-lived dusty discs. As
    such HD~62623 represents a missing link to the massive
    counterparts of these mass-losing stars, showing that a structured
    disc can also exist around such a highly luminous evolved star, and that this disc originates from the presence of a companion star.  

A different mechanism might lead to the
    formation of flattened dusty structures around other massive
    stars, but this finding definitely opens a promising route of interpretation of
    highly asymmetrical environments around massive stars in the
    frequent cases in which a low mass companion is suspected, albeit
    hardly detectable.

    Detailed physical modeling of the AMBER and MIDI data and a more
    complete study of the self-calibration algorithm applied to
    optical/IR interferometry will be presented in forthcoming
    papers. 

The presented self-calibration imaging algorithm practically demonstrates the potential of using the wavelength-differential phase in an image-reconstruction process. This self-calibration method is easily applicable to optical/infrared interferometric data and improves greatly the quality of the reconstructed images (which are now astrometrically linked from wavelength-to-wavelength).

In addition, this imaging method opens a wealth of possibilities that were considered as unfeasible before this paper: interferometric "integral-field spectroscopy", potentially reliable images from 2-Telescopes datasets, etc. This clearly opens new fields of research both in the signal processing domain (image synthesis techniques) and in the astrophysical domain (image-synthesis on "faint" targets, where the number of combined telescopes will always be limited: YSOs and AGNs).

    \vspace{1cm}
    \begin{acknowledgements}

We thank E. Thi\'ebaut for providing the image-reconstruction
software MIRA, M. Kishimoto, L. Rolland, K. Ohnaka for fruitful discussions,
A. Merand, and S. Stefl for supporting the observations.
The Programme National de Physique Stellaire (PNPS), the
Institut National en Sciences de l'Univers (INSU), and the Max Planck
Institute for Radioastronomy (MPIfR) are acknowledged for their financial and observing time support.
This research has made use of services from the Centre de Donn\'ees de Strasbourg (CDS), from
the Jean-Marie Mariotti Centre (JMMC), and from the NASA Exoplanet Science Institute (NExScI) to prepare and interpret the observations.
    \end{acknowledgements}

\end{document}